\documentclass[conference]{IEEEtran}
\IEEEoverridecommandlockouts
\usepackage[noadjust]{cite}
\usepackage{amsmath, amssymb, amsfonts}
\usepackage{algorithmic}
\usepackage{graphicx}
\usepackage{textcomp}
\usepackage{xcolor}

\def\BibTeX{{\rm B\kern-.05em{\sc i\kern-.025em b}\kern-.08em
    T\kern-.1667em\lower.7ex\hbox{E}\kern-.125emX}}
\begin{document}

\title{Frequency Modulation Aggregation\\ for Federated Learning
\thanks{This work is part of the project IRENE (PID2020-115323RB-C31), funded by MCIN/AEI/10.13039/501100011033 and supported by the Catalan government through the project SGR-Cat 2021-01207.}
}

\author{\IEEEauthorblockN{Marc Martinez-Gost\IEEEauthorrefmark{1}\IEEEauthorrefmark{2}, Ana Pérez-Neira\IEEEauthorrefmark{1}\IEEEauthorrefmark{2}\IEEEauthorrefmark{3}, Miguel Ángel Lagunas\IEEEauthorrefmark{2}}
\IEEEauthorblockA{
\IEEEauthorrefmark{1}Centre Tecnològic de Telecomunicacions de Catalunya, Spain\\
\IEEEauthorrefmark{2}Dept. of Signal Theory and Communications, Universitat Politècnica de Catalunya, Spain\\
\IEEEauthorrefmark{3}ICREA Acadèmia, Spain\\
\{mmartinez, aperez, malagunas\}@cttc.es
}}

\maketitle

\begin{abstract}
Federated edge learning (FEEL) is a framework for training models in a distributed fashion using edge devices and a server that coordinates the learning process. In FEEL, edge devices periodically transmit model parameters to the server, which aggregates them to generate a global model.
To reduce the burden of transmitting high-dimensional data by many edge devices, a broadband analog transmission scheme has been proposed. The devices transmit the parameters simultaneously using a linear analog modulation, which are aggregated by the superposition nature of the wireless medium.
However, linear analog modulations incur in an excessive power consumption for edge devices and are not suitable for current digital wireless systems.
To overcome this issue, in this paper we propose a digital frequency broadband aggregation. The scheme integrates a Multiple Frequency Shift Keying (MFSK) at the transmitters and a type-based multiple access (TBMA) at the receiver. Using concurrent transmission, the server can recover the type (i.e., a histogram) of the transmitted parameters and compute any aggregation function to generate a shared global model.
We provide an extensive analysis of the communication scheme in an additive white Gaussian noise (AWGN) channel and compare it with linear analog modulations. Our experimental results show that the proposed scheme achieves no drop in performance up to $-10$ dB and outperforms the analog counterparts, while requiring 14 dB less in peak-to-average power ratio (PAPR) than linear analog modulations.
\end{abstract}

\section{Introduction}

Edge learning is a type of machine learning that involves data processing and analysis directly on the edge devices that collect the data, rather than sending the data to a centralized server for processing. The main motivation behind edge learning is to reduce the amount of data 
transmitted to a central server, which can improve the efficiency and speed of data processing and reduce the cost of data transmission. 
Edge learning typically involves the use of lightweight machine learning algorithms that can run efficiently on edge devices with limited computational resources. One well-known framework is Federated Edge Learning (FEEL), in which a set of network devices train a shared model in a collaborative fashion \cite{mcmahan17, gaf22}. Each device performs local computation on its own data to update the model, and then shares the model parameters with the server. The latter generates a global model merging (e.g., averaging) the received parameters, which is then distributed back to the devices for further updates. FEEL implements a distributed version of stochastic gradient descent (SGD) which is executed iteratively until convergence. Furthermore, FEEL also addresses privacy and security concerns associated with transmitting sensitive data to third-party servers, since no raw data is shared.

One of the key challenges in FEEL is dealing with limited communication resources (e.g., limited power) at the edge devices, which can make it difficult to transmit large amounts of data over the network. This can lead to exhaustive delays, impacting the overall efficiency of the learning process. Since the server is not interested in recovering the individual parameters, one potential solution is over-the-air
computing (AirComp), that exploits the waveform superposition property of the wireless medium to support simultaneous transmission by several devices \cite{Nazer2007}. 
In \cite{Zhu19} the authors propose an broadband analog aggregation (BAA) technique in which each frequency resource of Orthogonal Frequency-Division Multiplexing (OFDM) is assigned to a model parameter. Each device modulates the information in the amplitude of the symbol and AirComp can be exploited for simultaneous transmission. However, relying on uncoded linear analog modulations (e.g., double-sideband modulation, DSB) may be unsuitable due to hardware concerns at the transmitter side: adjusting the power of the carrier due to variations in the modulated information can push the amplifier outside its operational range. Also, edge devices are usually energy constrained and may not be able to provide the transmission power that linear analog modulations require.
In this respect, \cite{zhu20} proposes a digital AirComp FEEL scheme in which only the sign of the gradients is transmitted using a binary frequency shift keying (BFSK). At the receiver side, the server resolves the aggregated gradient by majority voting. In \cite{sahin21}, the authors propose a FSK modulation for gradient majority voting that implements a non-coherent detector. Although the convergence of these schemes is empirically guaranteed at high signal-to-noise ratio (SNR) regime, the sign of the gradient may require many communication rounds at low SNR. Note also that these schemes rely on gradient-averaging, but cannot incorporate model-averaging.

To address the previous issues, in this work we propose the frequency modulation aggregation for FEEL.
Specifically, we generalize the work in \cite{zhu20, sahin21} to incorporate the magnitude of either the parameter or the gradient and speed up the learning process. We extend our previous work (\cite{Gost23, Gost23-2}), in which we proposed a Type-based Multiple Access (TBMA) scheme for estimation tasks, and apply it in a FEEL setting. TBMA lies in between linear analog aggregation and orthogonal multiple access, since resources (i.e., orthogonal waveforms) are assigned to orthogonal measurements, not devices. In other words, information is modulated according to their semantics \cite{Mergen2006}. By simultaneous transmission, the receiver recovers the type (i.e., histogram) of the transmitted data, over which the aggregation is computed. The waveform we propose is based on the Long Range (LoRa) modulation \cite{Chiani2019}, which consists in a Multiple Frequency Shift Key (MFSK) coupled with a chirp spread spectrum (CSS). This results in a waveform that is more robust to the noise channel and requires less power, which goes in line with the energy requirements of edge devices. On the other hand, CSS is also used to generate an Orthogonal Chirp Division Multiplexing (OCDM), so that several model parameters can be transmitted in the same time slot and reduce the time delay.
We limit the analysis to an additive white Gaussian noise (AWGN) channel and leave for future work the inclusion of fading channels. 

The remaining part of the paper proceeds as follows: Section II introduces the learning model and section III proposes the frequency digital aggregation model. Finally, Section IV complements the theoretical analysis with experimental results and Section V concludes the paper.

\section{Learning Model}
\label{learning_model}
\begin{figure}[t]
    \centering
    \includegraphics[width=\columnwidth]{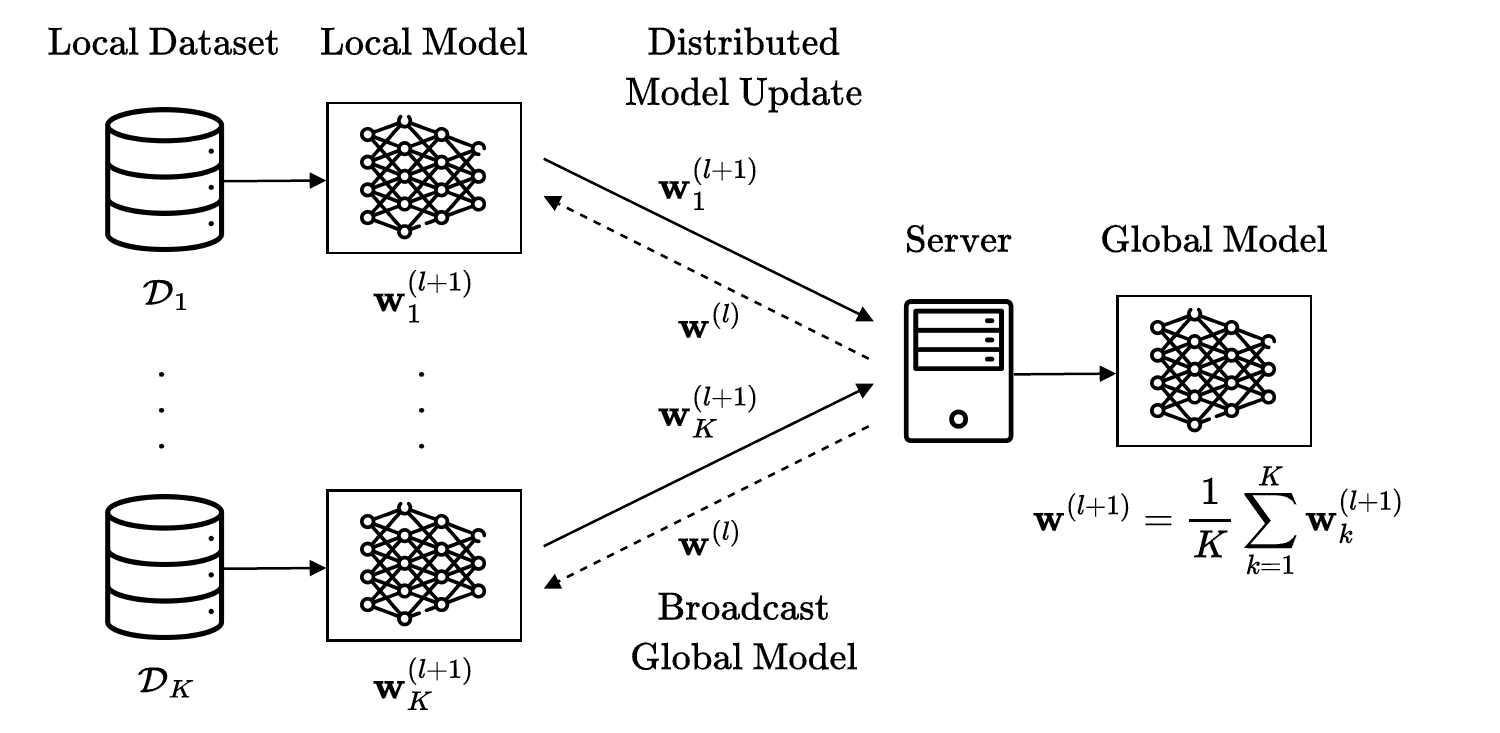}
  \caption{One communication round of a Federated Edge Learning scheme.}
  \label{fig: FEEL_scheme}
\end{figure}

Consider a FEEL system where a server coordinates the learning process among $K$ edge devices, as shown in Fig. \ref{fig: FEEL_scheme}.
Device $k=1,\dots,K$ collects a local dataset $\mathcal{D}_k$, consisting of $(\mathbf{x}_i,y_i)\in \mathcal{D}_k$, where the former is a $d$-sized vector training sample and the latter is the associated label. The learning model is represented by the parameter vector $\mathbf{w}$ of length $Q$, which is trained collaboratively across the network and orchestrated by the server. The \textit{local loss function} of the model $\mathbf{w}$ at device $k$ is
\begin{equation}
    F_k(\mathbf{w})=\frac{1}{|\mathcal{D}_k|} \sum_{(\mathbf{x}_i,y_i)\in \mathcal{D}_k} f(\mathbf{w},\mathbf{x}_i,y_i),
    \label{eq: local_loss}
\end{equation}
where $f(\mathbf{w},\mathbf{x}_i,y_i)$ is the sample loss function quantifying the prediction error of model $\mathbf{w}$ in dataset $\mathcal{D}_k$. For convenience, we assume equal size datasets $|\mathcal{D}_k|=D, \forall k$, and rewrite $f(\mathbf{w},\mathbf{x}_i,y_i)$ as $f(\mathbf{w})$. The \textit{global loss function} on all the distributed datasets is
\begin{equation}
    F(\mathbf{w})=\frac{\sum_{k=1}^K \sum_{i\in\mathcal{D}_k}f(\mathbf{w})}{\sum_{k=1}^K |\mathcal{D}_k|}=
    \frac{1}{K}\sum_{k=1}^K F_k(\mathbf{w}).
    \label{eq: global_loss}
\end{equation}

The goal of the learning process is to find the model $\mathbf{w}^*$ that minimizes the global loss function $F(\mathbf{w})$. In a FEEL setting, the learning process is iterative and split into communication rounds. At communication round $l$ the server broadcasts the global model $\mathbf{w}^{(l)}$ and each device updates it using its own local dataset. We assume local models are trained using SGD, this is,
\begin{equation}
    \mathbf{w}_k^{(l+1)} = \mathbf{w}^{(l)} - \eta\nabla F_k(\mathbf{w}^{(l)}),
    \label{eq: SGD}
\end{equation}
where $\eta$ is the step-size and $\nabla$ is the gradient operator. To conclude the communication round, the edge devices send their local models towards the server, which computes the global model by averaging them as
\begin{equation}
    \mathbf{w}^{(l+1)} = \sum_{k=1}^K \mathbf{w}_k^{(l+1)}
    \label{eq: global_update}
\end{equation}

\section{Communication Model}

Local models are transmitted over a broadband multiple access channel (MAC) towards the server. We assume AWGN channels with no fading, since the goal is to show the feasibility of deploying a FEEL system supported by a frequency-based aggregation scheme. Thus, we leave fading channels for future work.

We assume perfect time and phase synchronization to exploit AirComp accurately. This  can be achieved via standardized techniques, such as timing advance in LTE \cite{timing_advance}, or protocols as AirShare in which a clock is shared among sensors for coherent transmissions \cite{airshare}.

\subsection{Frequency Modulation for Aggregation}
Given the $k$-th local model at communication round $l$, $\mathbf{w}_k^{(l)}$,  each model parameter could be analog modulated in frequency as
\begin{equation}
z_{k,q}^{FM}[n] = A_c\sqrt{\frac{2}{N}} \cos\left(\frac{\pi(2\mathbf{w}_k^{(l)}[q]+1)}{2N}n\right),
\label{eq:dct_tbma_no}
\end{equation}
for $n=0,\dots,N-1$, where $\mathbf{w}_k^{(l)}[q]$ is the $q$-th entry of vector $\mathbf{w}_k^{(l)}$. However, the demodulation process is more complex and does not fully exploits the AirComp framework. Thus, the model parameters will be quantized, mapped into one of the available discrete indexes $m\in[0,N-1]$, and digitally modulated via MFSK. Since we assume a uniform quantizer $\mathcal{Q}(\cdot)$ and a linear mapping $g(\cdot)$, every model parameter can be exactly recovered up to its quantization error. 

We define the $q$-th quantized model parameter at device $k$ as
\begin{equation}
    m_{k,q}=g\left(\mathcal{Q}\left(\mathbf{w}_k^{(l)}[q]\right)\right)\in [0,\dots,N-1],
    \label{eq: quantizer}
\end{equation}
in which we have dropped the superscript $l$ without loss of generality. This is modulated as
\begin{equation}
z_{k,q}[n] = A_c\sqrt{\frac{2}{N}} \cos\left(\frac{\pi(2m_{k,q}+1)}{2N}n\right),
\label{eq:dct_tbma}
\end{equation}
for $n=0,\dots,N-1$. Notice that \eqref{eq:dct_tbma} corresponds to the LoRa modulation (see \cite{Chiani2019}) using the Discrete Cosine Transform (DCT) and not the Discrete Fourier Transform (DFT) basis. Thus, computing the inverse DCT of \eqref{eq:dct_tbma} yields a peak of amplitude $A_c$ located at frequency $m_{k,q}$ (see \cite{Gost23} for a detailed explanation of this modulation and its benefits). Notice that this demodulation step is simpler than in the analog case.

In \eqref{eq:dct_tbma} frequency resources are assigned to orthogonal parameters, not devices. This is the core of TBMA, a semantic-aware multiple access scheme that lies between pure AirComp and orthogonal multiple access. In TBMA, AirComp occurs when two users modulate the same information ($m_{k,q}=m_{k',q}, k\neq k'$), as they generate the same waveform and the channel provides a constructive interference. By having a bank of matched filters, the receiver recovers a histogram of the parameters.

At a given time slot all the devices communicate simultaneously their corresponding parameter $\mathbf{w}_k^{(l)}[q]$. After downconversion, filtering and sampling, the signal at the input of the receiver is
\begin{equation}
    y_q[n] = \sum_{k=1}^K z_{k,q}[n] + w[n],
    \label{eq:tbma}
\end{equation}
where $w$ corresponds to AWGN samples. Computing the matched filter (i.e., DCT) of $y_q$ and normalizing by $1/A_cK$ results in
\begin{equation}
    \textbf{r}_q = \frac{1}{K} [K_0,\dots,K_{N-1}]^T + [\tilde{w}_0,\dots,\tilde{w}_{N-1}]^T = \tilde{\textbf{p}}_q + \tilde{\textbf{w}}, 
    \label{eq:empirical_measure}
\end{equation}
where the $n$-th entry in $\tilde{\textbf{p}}_q$ corresponds to the fraction of devices $K_m/K$ that transmitted the parameter $m\in[0,N-1]$ and $\tilde{w}_n$ is Gaussian noise with power $\sigma^2 = N_0/2A_c^2K^2$. Provided that $K>N$, which is reasonable in a FEEL scenario,  $\textbf{r}_q$ corresponds to a nosy version of the empirical measure $\tilde{\textbf{p}}_q$, this is, a histogram of $m_{k,q}$, with respect to $q$.

Finally, the $q$-th entry of \eqref{eq: global_update}, this is, the $q$-th global parameter, is obtained by taking the mean of the noisy empirical measure and mapping the result back to the original range. Since $g(\cdot)$ is a linear transformation, the existence of $g^{-1}(\cdot)$ is guaranteed. This results in
\begin{align}
    \mathbf{w}^{(l)}[q] =&\
    g^{-1}\left(\sum_{n=0}^{N-1} n \mathbf{r}_q[n]\right)\nonumber\\ \nonumber=&\
    g^{-1}\left(\frac{1}{K}\sum_{k=1}^K m_{k,q} + \sum_{n=0}^{N-1}w[n]\right)\\=&\
    \frac{1}{K}\sum_{k=1}^K \mathcal{Q}\left(\mathbf{w}_k^{(l)}[q]\right) + w_{eq},
    \label{eq: parameter_mean}
\end{align}
where $w_{eq}$ is the equivalent noise sample with power $\sigma_{eq}^2 = N_0N/2A_c^2K^2$. Notice that the effect of adopting a digital modulation causes an additional source of error associated to the quantization noise. While in \eqref{eq: parameter_mean} we proceed with the standard aggregation function (i.e., mean), TBMA allows to compute any other aggregation function. In this way, TBMA generalizes AirComp to other edge-learning architectures \cite{zhu20-2}.


While the global model update in \eqref{eq: global_update} hints that FEEL preserves privacy because no local data leaves the device, the aggregation in \eqref{eq: parameter_mean} shows that AirComp further preserves privacy because th server has no access to the individual parameter models.

\subsection{Orthogonal Chirp Division Multiplexing}
In our previous work (\cite{Gost23, Gost23-2}) we motivated the benefits of \eqref{eq:dct_tbma} in integrating a chirp spread spectrum (CSS) as in LoRa. Spreading a narrowband signal over a wide bandwidth supports the development of long range communications while severely improving the receiver sensitivity by 20 dB. LoRa has gained relevance in the context of Internet of Things (IoT) and machine to machine (M2M) communications either terrestrial or via satellite, because the modulation allows to reduce the energy consumption and diminish the effect of interferences to other devices. Nonetheless, in this section we also exploit the chirp to multiplex several waveforms in the frequency domain.

The size $Q$ of current learning models is very large (even in the order of millions), meaning that establishing a time division multiple access with respect to $Q$ would induce extremely large delays.
This justifies that the existing AirComp methods for FEEL rely on OFDM: the parameter $\mathbf{w}_k[q]$ is modulated in the amplitude of the carrier, and different frequencies are allocated at different $q$. 
However, in the presented scheme this is not possible because several users may transmit at different frequencies and a single AirComp transmission occupies beyond a single OFDM frequency carrier. Conversely, we propose Orthogonal Chirp Division Multiplexing (OCDM) to generate $P$ orthogonal chirps so that $P$ different parameters can be transmitted simultaneously and reduce the delay.

The authors in \cite{ouy16} show that the following digital chirp,
\begin{equation}
    \psi_p[n] = e^{-j\frac{\pi}{N}\left(n-p\right)^2},\quad n=0,\dots,N-1,
\end{equation}
allows to generate $P$ orthogonal chirps for $p=0,\dots,P-1$ and for an even $N$. This assumption holds since $N$ comes from the quantization step and it is usually a power of 2. Thus, the discrete OCDM signal at device $k$ and parameter $\mathbf{w}_k[q]$ is
\begin{equation}
    s_{k,q}[n] = z_{k,q}[n] e^{-j\frac{\pi}{N}\left(n-q\right)^2},\quad n=0,\dots,N-1.
    \label{eq: ocdm_signal}
\end{equation}

\begin{figure*}[t]
    \centering
    \includegraphics[width=\textwidth]{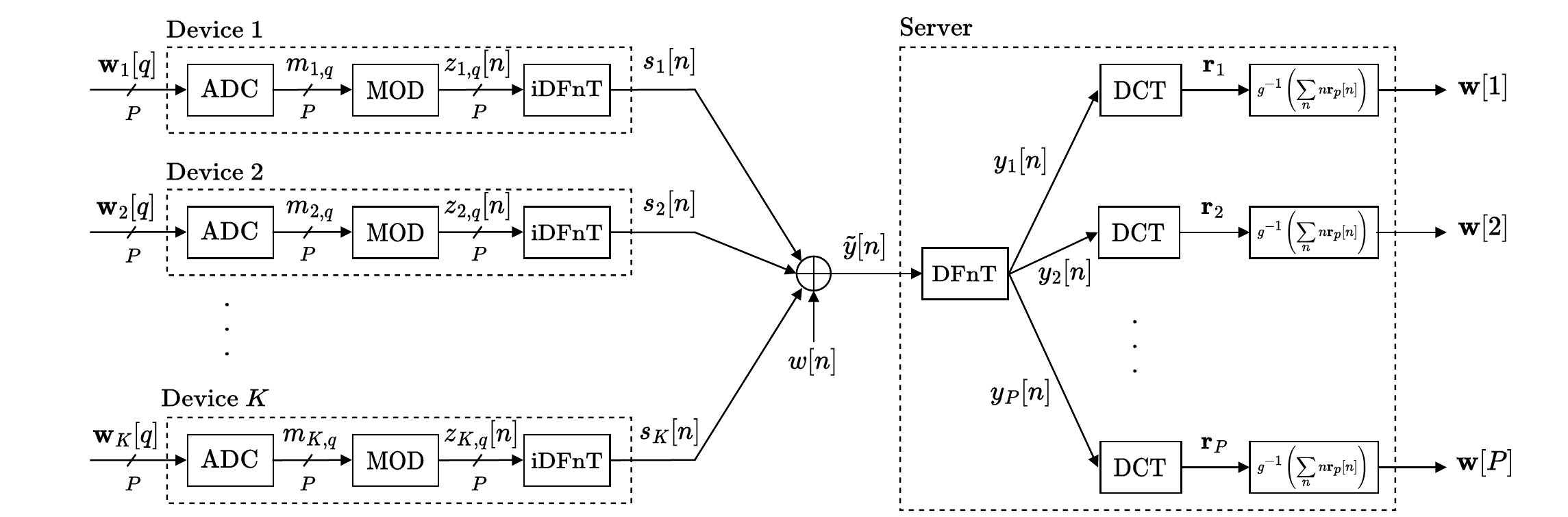}
  \caption{Transmitter and receiver designs for the frequency modulation-based aggreation scheme for FEEL.}
  \label{fig: communication_scheme}
\end{figure*}

Each device can generate $P$ chirps in parallel and transmit them simultaneously. Notice that these chirps are identical across devices for a same parameter index $q$.
Since $\psi_p$ and $\psi_p'$ for $p\neq p'$ are orthogonal, at the receiver side the different indexes $q$ can be recovered using a bank of matched filters with $\psi_p$ for $p=0,\dots,P-1$.

The authors show that OCDM corresponds to the inverse Discrete Fresnel Transform (iDFnT), which is a linear operator. The synthesis of a bank of discretized modulated chirp waveforms can be realized by the IDFnT as
\begin{equation}
    \mathbf{s}=\mathbf{\Phi} \mathbf{z},
    \label{eq: iDFnT}
\end{equation}
where $\mathbf{s}$ and $\mathbf{z}$ are the vectorized version of $s[n]$ and $z[n]$, and
\begin{equation}
    \mathbf{\Phi}(n,q) = \frac{1}{\sqrt{N}}e^{-j\frac{\pi}{4}}e^{-j\frac{\pi}{N}\left(n-q\right)^2}.
\end{equation}

Similarly, the bank of matched filters at the receiver can be implemented with the DFnT as
\begin{equation}
    \mathbf{z}=\mathbf{\Phi}^H \mathbf{s}
    \label{eq: DFnT}
\end{equation}

The number of time slots required to transmit all the parameters is
\begin{equation}
    N_s = \left\lceil\dfrac{Q}{P}\right\rceil,
\end{equation}
where $\lceil\cdot\rceil$ is the ceil function and it represents a reduction in time delay by a factor of $Q/N_s$.
The reduction in the time by transmitting a set of $P$ parameters simultaneously comes at the expenses of increasing the transmission bandwidth. The OCDM signal in \eqref{eq: ocdm_signal} increases the bandwidth proportional to the number of orthogonal chirps $P$.

\subsection{Frequency-based Aggregation System}

Figure \ref{fig: communication_scheme} shows the overall communication scheme based on the frequency modulation aggregation for a FEEL system. The analog-to-digital converter (ADC) corresponds to \eqref{eq: quantizer}, and arrows with subscript $P$ indicate that there are $P$ branches executed in parallel. The standard communication blocks (e.g., downconversion, filtering, etc.) are omitted for simplicity. The transmitted signal by device $k$ corresponds to
\begin{equation}
    s_k[n] = 
    \sum_{p=0}^{P-1}s_{k,p}[n]
\end{equation}
and the received signal at the input of the receiver is $\tilde{y}[n]$. Notice that the receiver needs to wait until $N$ time samples are received to demodulate.

\subsection{SNR guarantees of frequency-based aggregation}
\label{sec: SNR}
Because the frequency-based aggregation scheme uses a constant envelope waveform, the transmission power is constant and the amplitude of all devices can be set to meet a specific SNR requirement. Conversely, consider the following DSB waveform,
\begin{equation}
    d_{k,q}[n] = A_qm_{k,q}\sqrt{\frac{2}{N}} \cos\left(\frac{2\pi F_q}{N}n\right)
    \label{eq: DSB}
\end{equation}
with discrete frequency $F_q$, that corresponds to the BAA model in \cite{Zhu19}. In order to preserve the AirComp aggregation using linear analog modulations it is not possible for all devices to transmit at the same SNR, since the transmitted power depends on the data $m_{k,q}$. 

To conduct fair comparisons between both modulations, in Sec. \ref{sec:experiments} we will operate at average SNR. The amplitude $A_q$ in \eqref{eq: DSB} will be set so that the average power transmitted per device at index $q$ guarantees the SNR level. However, notice that implementing this approach is not feasible, since it requires previous knowledge of the data transmitted by all devices.

Alternatively, the SNR in the frequency-based aggregation is completely determined by the amplitude $A_c$ in \eqref{eq:dct_tbma}.
This represents an advantage of frequency modulations with respect to amplitude modulations for distributed AirComp systems: DSB cannot guarantee a certain SNR level, because it requires prior knowledge of all the transmitted data. On the other hand, since frequency modulated waveforms are designed to have unit power, the SNR can be achieved effortlessly.

\begin{figure*}[t!]
    \centering
    \includegraphics[width=0.9\textwidth]{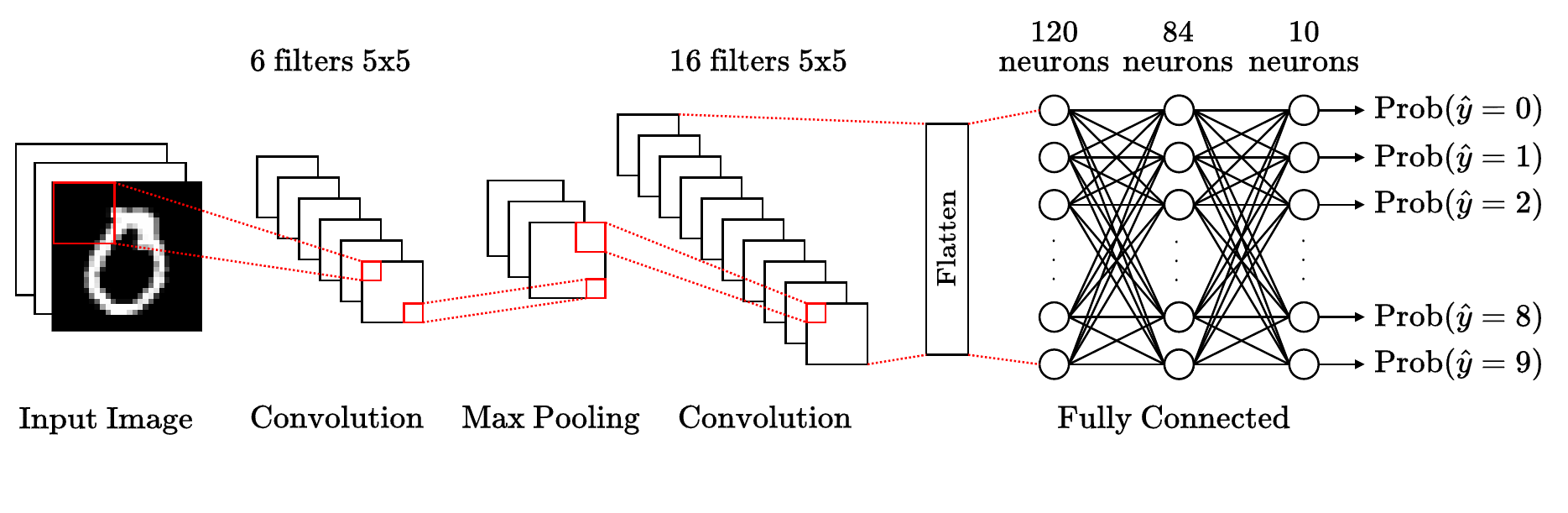}
  \caption{Architecture of the CNN model used to train the FEEL system for a classification task. The output is the probability that the input image belongs to each class.}
  \label{fig: FEEL_CNN}
\end{figure*}

\subsection{Robustness Against Adversarial Attacks by OCDM}

In the BAA model of \cite{Zhu19}, the authors integrate a direct sequence spread spectrum (DSSS) technique to
encode the model parameters before transmission. While the spreading code does not provide any benefit in terms of SNR, it reduces the interference created to neighboring devices and provides robustness against adversarial attacks. Any attacker will not be able to reach the server unless it knows the specific pseudorandom-noise code associated to every parameter.

The proposed scheme in this paper implements a spreading technique to multiplex several parameter updates in the frequency domain. Thus, it inherits the benefits of spreading the signal with respect to adversarial attacks.
If the assignment between the columns of  $\mathbf{\Phi}$ and indexes $q$ is unknown to any device outside the network, the de-spreading at the receiver side will reduce any attack to the noise level. Similarly, the system requires tight time synchronization for both OCDM to work properly. While this sophisticates the communication system, it also mitigates the effect any adversarial attack can create over the learning process.

\section{Performance Evaluation}
\label{sec:experiments}

\subsection{Experimental setup}
We consider the deployment of a FEEL system in which a server coordinates the learning process of $K=50$ devices. The task is image classification using the MNIST dataset \cite{mnist}, consisting in black and white images of handwritten digits ranging from 0 to 9. The FEEL system is deployed using Flower \cite{flower}, a framework for training machine learning models in a federated fashion.

The learning model consists in a convolutional neural network (CNN), whose architecture is shown in Fig. \ref{fig: FEEL_CNN}. All the filters are followed by a ReLu activation function, except for the latter, which incorporates a softmax layer. Thus, the output is the probability that the predicted label $\hat y$ belongs to each class. This architecture is chosen as it is simple enough to solve the classification problem with high accuracy in a centralized setting. This allows to see the effect of including a communication layer in the learning process.
The parameter vector $\mathbf{w}_k$ corresponds to all the weights of the $k$-th CNN stacked sequentially. The local loss function used to train each CNN is cross-entropy, which is backpropagated using the Adam optimizer. Each device runs one optimization step per communication round and the overall FEEL system is trained for 10 communication rounds. The performance is measured by the average test accuracy (i.e., percentage of test samples correctly classified) across all devices at each communication round.

The ultimate goal of the experiments is to validate the feasibility of deploying a frequency-based aggregation for FEEL and evaluate the performance of the modulation in an AWGN channel. Thus, we assume the datasets $\mathcal{D}_k$ are independent and identically distributed, and all $K$ devices are used for training and evaluation. The proposed frequency scheme is trained with $N=\{32, 256\}$ samples per parameter. The magnitude of the parameters is clipped to 0.5, corresponding to the first and last indexes of \eqref{eq: quantizer}. This parameter is set after training a centralized CNN and observing the distribution of parameters. Clipping the parameters to 0.5 in magnitude affects less than $1\%$ of them.
The chosen benchmark is BAA, this is, a linear-analog scheme using the DSB modulation and pure AirComp aggregation (see \eqref{eq: DSB}). For a fair comparison, we also clip the parameters to 0.5 and also set parameters below $4\cdot10^{-3}$ to zero, which would incur in excessive power consumption. The last lower bound comes from the step size of the quantizer in MFSK for $N=256$, which corresponds to the smallest number that the quantizer can resolve.

In the literature the effect of the SNR in the convergence of the algorithm is understudied and a high SNR regime (e.g., 10 dB) is usually assumed. As mentioned in Sec. \ref{sec: SNR}, the experiments are conducted for average transmitted power (i.e., average SNR). For DSB, at communication round $l$ and parameter index $q$, we set the amplitude according to
\begin{equation}
    A_q = \sqrt{\frac{P_T}{\sum_{k=1}^K \left(\mathbf{w}_{k,q}^{(l)}\right)^2}},
\end{equation}
where $P_T$ is the average transmitted power and the noise power is assumed to be 0 dB. Regarding the frequency modulation, \eqref{eq:dct_tbma}, we set $A_c=\sqrt{P_T}, \forall q, \forall k$. Furthermore, we assume that all devices have enough transmission power.

\subsection{Numerical Results}
\begin{figure}[t]
\centering
\includegraphics[width=\columnwidth]{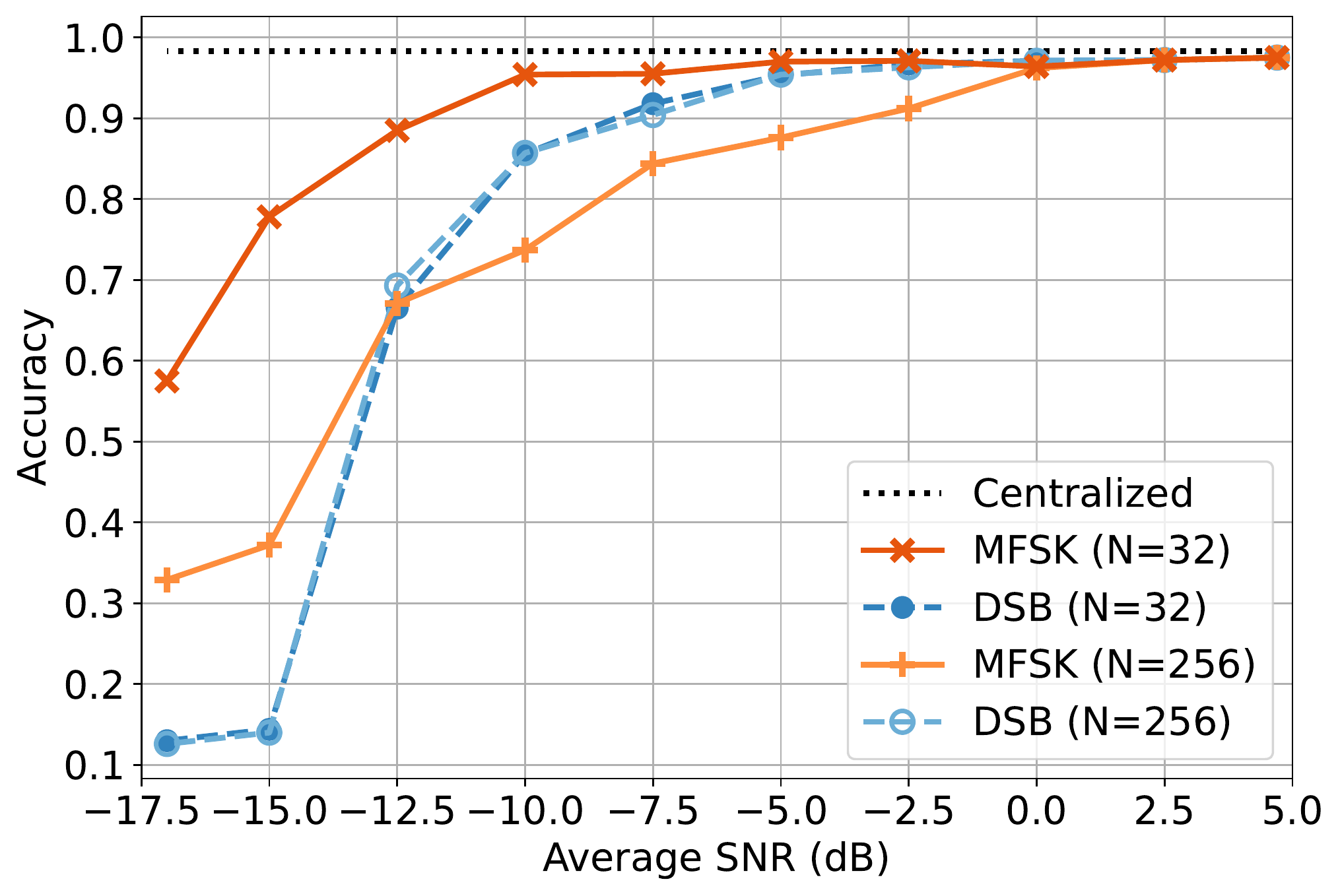}
\caption{Accuracy of the FEEL system for the digital frequency modulation (MFSK) and the linear analog modulation (DSB) for $N=\{32,256\}$, along with the upper bound of a centralized trained model.}
\label{fig:sim_accuracy_SNR}
\vspace{-0 pt}
\end{figure}

A performance comparison between transmission using linear analog modulations (namely, BAA), and the proposed frequency modulations is presented in Fig. \ref{fig:sim_accuracy_SNR} for $N=\{32,256\}$. The performance of a centralized model with no FEEL architecture and no communication scheme is presented as well as an upper bound on the model performance ($98\%$).

As expected, the performance of MFSK exhibits a trade-off between the number of users $K$ and the number of resources $N$: When $N=32$, $K>N$ and the receiver recovers an accurate approximation of the empirical measure. In this case, the performance of the frequency modulation remains almost intact up to -10 dB, which does not happen for the DSB scheme. Conversely, when $N=256$, $K<N$ and the benefits of TBMA cannot be exploited. The performance of the frequency modulation degrades rapidly. In conclusion, the number of frequency bins $N$ in MFSK can be tuned between 2 (i.e., BFSK in sign-SGD) and the maximum number of samples. On the other hand, the accuracy of DSB is independent of $N$. 

To assess the power both modulations require, we compute the peak-to-average power ratio (PAPR) across the $Q$ parameters, which helps in determining the power requirements of each communication system. While in MFSK, the PAPR is 0 dB (see Sec. \ref{sec: SNR}), DSB achieves 14 dB. This indicates that DSB is less power efficiency, which is undesirable for edge devices (e.g., IoT networks). Moreover, a high PAPR signal may be more susceptible to distortion and interference, which can affect the performance of the system.

\section{Conclusions}

We have proposed a new aggregation scheme for FEEL based on a frequency modulation.
The digital modulation, MFSK, resorts on TBMA to assign frequency resources and AirComp can be exploited through simultaneous transmission. After demodulation the receiver obtains a histogram of the transmitted parameters, which allows to compute any aggregation function. Furthermore, we have proposed OCDM to multiplex several parameters in frequency, which comes for free as the original waveform can implement the LoRa modulation. In conclusion, the overall communication scheme is simple to implement at each transmitter, since it only requires an ADC and an MFSK modulator. The experiments, conducted in an AWGN channel, show that the FEEL scheme inherits the benefits from a frequency modulation and experiences no drop in learning performance up to $-10$ dB. Moreover, the proposed scheme requires up to 14 dB less power in terms of PAPR.
In future work we expect to study the performance of the scheme with respect to different quantization levels $N$, and propose different aggregation functions that can mitigate the effect of channel fading and accelerate the learning convergence.


\bibliographystyle{IEEEbib}
\bibliography{refs}

\begin{thebibliography}{10}

\bibitem{mcmahan17}
Brendan McMahan, Eider Moore, Daniel Ramage, Seth Hampson, and Blaise~Aguera
  y~Arcas,
\newblock ``Communication-efficient learning of deep networks from
  decentralized data,''
\newblock in {\em Artificial intelligence and statistics}. PMLR, 2017, pp.
  1273--1282.

\bibitem{gaf22}
Tomer Gafni, Nir Shlezinger, Kobi Cohen, Yonina~C Eldar, and H~Vincent Poor,
\newblock ``Federated learning: A signal processing perspective,''
\newblock {\em IEEE Signal Processing Magazine}, vol. 39, no. 3, pp. 14--41,
  2022.

\bibitem{Nazer2007}
Bobak Nazer and Michael Gastpar,
\newblock ``Computation over multiple-access channels,''
\newblock {\em IEEE Transactions on Information Theory}, vol. 53, no. 10, pp.
  3498--3516, 2007.

\bibitem{Zhu19}
Guangxu Zhu, Yong Wang, and Kaibin Huang,
\newblock ``Broadband analog aggregation for low-latency federated edge
  learning,''
\newblock {\em IEEE Transactions on Wireless Communications}, vol. 19, no. 1,
  pp. 491--506, 2019.

\bibitem{zhu20}
Guangxu Zhu, Yuqing Du, Deniz G{\"u}nd{\"u}z, and Kaibin Huang,
\newblock ``One-bit over-the-air aggregation for communication-efficient
  federated edge learning: Design and convergence analysis,''
\newblock {\em IEEE Transactions on Wireless Communications}, vol. 20, no. 3,
  pp. 2120--2135, 2020.

\bibitem{sahin21}
Alphan Şahin, Bryson Everette, and Safi~Shams Muhtasimul~Hoque,
\newblock ``Distributed learning over a wireless network with {FSK}-based
  majority vote,''
\newblock in {\em 2021 4th International Conference on Advanced Communication
  Technologies and Networking (CommNet)}, 2021, pp. 1--9.

\bibitem{Gost23}
Marc~M. Gost, Ana Pérez-Neira, and Miguel~Ángel Lagunas,
\newblock ``{DCT-based air interface design for function computation},''
\newblock {\em IEEE Open Journal of Signal Processing}, pp. 1--9, 2023.

\bibitem{Gost23-2}
Marc Martinez-Gost, Ana P{\'e}rez-Neira, and Miguel~{\'A}ngel Lagunas,
\newblock ``{LoRa}-based over-the-air computing for sat-{IoT},''
\newblock {\em arXiv preprint arXiv:2306.16333}, 2023.

\bibitem{Mergen2006}
G.~Mergen and L.~Tong,
\newblock ``Type based estimation over multiaccess channels,''
\newblock {\em IEEE Transactions on Signal Processing}, vol. 54, no. 2, pp.
  613--626, 2006.

\bibitem{Chiani2019}
Marco Chiani and Ahmed Elzanaty,
\newblock ``On the {LoRa} modulation for {IoT}: Waveform properties and
  spectral analysis,''
\newblock {\em IEEE Internet of Things Journal}, vol. 6, no. 5, pp. 8463--8470,
  2019.

\bibitem{timing_advance}
TS~36.213,
\newblock ``Physical layer procedures,''
\newblock Tech. {R}ep., 3GPP, v.12.6.0, July 2015.

\bibitem{airshare}
Omid Abari, Hariharan Rahul, Dina Katabi, and Mondira Pant,
\newblock ``Airshare: Distributed coherent transmission made seamless,''
\newblock in {\em 2015 IEEE Conference on Computer Communications (INFOCOM)},
  2015, pp. 1742--1750.

\bibitem{zhu20-2}
Guangxu Zhu, Dongzhu Liu, Yuqing Du, Changsheng You, Jun Zhang, and Kaibin
  Huang,
\newblock ``Toward an intelligent edge: Wireless communication meets machine
  learning,''
\newblock {\em IEEE Communications Magazine}, vol. 58, no. 1, pp. 19--25, 2020.

\bibitem{ouy16}
Xing Ouyang and Jian Zhao,
\newblock ``Orthogonal chirp division multiplexing,''
\newblock {\em IEEE Transactions on Communications}, vol. 64, no. 9, pp.
  3946--3957, 2016.

\bibitem{mnist}
Li~Deng,
\newblock ``The {MNIST} database of handwritten digit images for machine
  learning research,''
\newblock {\em IEEE Signal Processing Magazine}, vol. 29, no. 6, pp. 141--142,
  2012.

\bibitem{flower}
Daniel~J. Beutel, Taner Topal, Akhil Mathur, Xinchi Qiu, Javier
  Fernandez-Marques, Yan Gao, Lorenzo Sani, Kwing~Hei Li, Titouan Parcollet,
  Pedro Porto~Buarque de~Gusm{\~a}o, and Nicholas~D. Lane,
\newblock ``Flower: A friendly federated learning framework,''
\newblock Open-Source, mobile-friendly Federated Learning framework, Mar. 2022.

\end{thebibliography}

\end{document}